\documentclass[titlepage,oneside,12pt]{article}
\usepackage[tiny,rm]{titlesec}
\newpagestyle{trbstyle}{
\sethead{Polson and Sokolov}{}{\thepage}
}
\titleformat{\section}{\bfseries}{}{0pt}{\uppercase}
\titlespacing*{\section}{0pt}{12pt}{*0}
\titleformat{\subsection}{\bfseries}{}{0pt}{}
\titlespacing*{\subsection}{0pt}{12pt}{*0}
\titleformat{\subsubsection}{\itshape}{}{0pt}{}
\titlespacing*{\subsubsection}{0pt}{12pt}{*0}

\usepackage{enumitem}
\setlist[1]{labelindent=0.5in,leftmargin=*}
\setlist[2]{labelindent=0in,leftmargin=*}

\usepackage{ccaption}
\usepackage{amsmath}
\makeatletter
\renewcommand{\fnum@figure}{\textbf{FIGURE~\thefigure} }
\renewcommand{\fnum@table}{\textbf{TABLE~\thetable} }
\makeatother
\captiontitlefont{\bfseries \boldmath}
\captiondelim{\;}

\usepackage{times}

\usepackage[T1]{fontenc}
\usepackage[urw-garamond]{mathdesign}

\usepackage[sort,numbers]{natbib}
\newcommand{\trbcite}[1]{\citeauthor{#1} ({\it \citenum{#1}})}
\setcitestyle{round}

\usepackage{totcount}
	\regtotcounter{table} 	
	\regtotcounter{figure} 	

\usepackage{amsmath}

\usepackage{amsfonts}
\usepackage{hyperref}
\usepackage{float}
\usepackage{amsthm,amsmath}
\usepackage{enumerate}
\usepackage{graphicx}

\begin{document}

\title{\bf Bayesian Particle Tracking of Traffic Flows}
  \author{Nicholas Polson\hspace{.2cm}\\
    Booth School of Business, University of Chicago\\
    and \\
    Vadim Sokolov \\
    Argonne National Laboratory\\
    }
\thispagestyle{empty}
\maketitle

\newpage

\thispagestyle{empty}
\section{Abstract}
We develop a Bayesian particle filter for tracking traffic flows that is capable of capturing non-linearities and discontinuities present in flow dynamics. Our  model includes a hidden state variable that captures sudden regime shifts between traffic free flow, breakdown and recovery. We develop  an efficient particle learning algorithm for real time on-line inference of states and parameters. This requires a two step approach, first, resampling the current particles, with a mixture predictive distribution and second, propagation of states using the conditional posterior distribution. Particle learning of parameters follows from updating recursions for conditional sufficient statistics. To illustrate our methodology, we analyze measurements of daily traffic flow from the Illinois interstate I-55 highway system. We demonstrate how our filter can be used to inference the change of traffic flow regime on a highway road segment based on a measurement from freeway single-loop detectors. Finally, we conclude with directions for future research.
\newpage

\section{Introduction}
Modeling traffic dynamics for a transportation network with highways, arterial roads and public transit is an important task for effectively managing traffic flow. A major goal is to provide traffic flow conditions on highways from field measurements fusing in-ground loop detectors or GPS probes. Traffic managers make decisions based on model forecasts to regulate ramp metering, apply speed harmonization, or change road pricing as congestion mitigation strategies. The general public uses model-based  predictions for assessing departure times, and travel route choices, among other factors.

We propose a dynamic state-space model which incorporates a latent switching  variable together with a traffic flow state variable to capture the non-linearities and discontinuities in traffic patterns. Our dynamic  model allows for three traffic flow regimes: free flow, breakdown and recovery.  A physical interpretation of the change in the flow regime is a traffic queue with congestion inside the queue and free flow outside. In addition to switching variable and  traffic flow states, we track a state that measures the rate of degradation or recovery. The challenge is to filter traffic flow measurements which are sparse, from both fixed  and moving sensors which are located at a small number of locations at any given point in time. Statistical inference is further complicated by noisy non-Gaussian observations; for example, \trbcite{ras:03}  show that the data generating distribution for video cameras is a mixture of Poissons.

In order to achieve real-time sequential inference, we develop an efficient particle filter and learning algorithm. This allows us to track  the speed of traffic flow together with the latent states that describe regime switches, and the rate of degradation (recovery).  To illustrate our methodology, we model a data from a network of in-ground loop detectors on Chicago's Interstate I-55 with measurements on  speed, counts and occupancy of traffic flow. The sequential nature of particle filtering makes frequent updating feasible, and therefore, our method provides an alternative to Markov Chain Monte Carlo (MCMC) simulation methods. The computational cost of the latter grows linearly with the number of observations. We build on particle filters for  traffic flow problems (see \trbcite{mih07}) who use the evolution dynamics as a proposal distribution  before re-sampling,  the so-called  bootstrap or sampling/importance resampling (SIR) filter. We improve the filtering efficiency by developing a Rao-Blackwellized which is also flexible enough to incorporate particle learning.

Our approach builds upon existing statistical methods in a number of ways. \trbcite{tebaldi1998bayesian}  infer network route flows and \trbcite{westgate2013travel} develop  MCMC methods to estimate travel time reliability for ambulances using noisy GPS for both path travel time and individual road segment travel time distributions. \trbcite{anacleto2013multivariate} propose a dynamic Bayesian network to model external intervention techniques to accommodate situations with suddenly changing traffic variables. Another class of probabilistic methods rely on using image processing techniques for traffic estimation, for example in \trbcite{kamijo_traffic_2000} authors developed an algorithm that is based on Markov random field that allows analyzing images from intersections to detect flows and accidents. 

Small changes in consecutive speed measurements can be explained  by sensor noise but significant changes require estimation of probability of a flow regime switch, which is represented by a discontinuity in traffic flow speed.   Previous work on estimating traffic flows use extensions of the Kalman filter and  relies heavily on Gaussianity assumptions, see \trbcite{gaz71,sch10, wang05, work08}.  Recently, \trbcite{bla12, polson2014bayesian} showed that  dynamic properties of traffic flow such as discontinuities (or shock waves), lead to forecast distributions that are mixtures. Our non-Gaussian state-space model will explicitly capture such a behavior. 

Our approach builds on the current literature in a number of ways:
\begin{itemize}
	\item Provides a hierarchical model rather than a conservation law based model. Our approach avoids boundary condition estimation 
	\item A predictive likelihood particle filter provides an efficient estimation strategy. Our filter is less sensitive to measurement outliers and less prone to particle degeneracy
	\item Tracks traffic flow variables, such as traffic flow speed together with additional latent variables for regime switching and a degeneracy (recovery) rate.
\end{itemize}

One class of previously considered problems involves estimating (i)  turn counts on urban controlled intersections [\trbcite{lan_real-time_1999, ghods_real-time_2014}], (ii) flows or travel times on network routes [\trbcite{kwong_arterial_2009, wu_cellpath:_2015, rahmani_non-parametric_2015, nantes_real-time_2015, fowe_microstate_2013, dellorco_bee_2015}], (iii) travel times and densities on an individual road segment [\trbcite{coifman_speed_2009, zheng_urban_2013, zhan_urban_2013, seo_probe_2015, seo_estimation_2015, bachmann_comparative_2013}], and (iv) queue position on highways or arterial roads [\trbcite{vigos_real-time_2008, jeff_ban_real_2011, zhan_lane-based_2015, lee_real-time_2015}]. There are several types of algorithms, which can be categorized into the following groups (i) time series analysis, (ii) machine or deep learning, and (iii) model-based analysis in a form of state-space models. We tackle the latter, which requires a particle filtering algorithm to perform inference. 

A machine-learning algorithm  was provided by \trbcite{sun_bayesian_2006}, where authors proposed a Bayes network algorithm for forecasting traffic flow. The idea is to derive the conditional probability of a traffic state on a given road, given states on topological neighbors on a road network. The resulting joint probability distribution is a mixture of Gaussians. Bayes networks for estimating travel times were suggested earlier by \trbcite{horvitz2012prediction}. This approach eventually became a commercial product that led to the start of Inrix, a traffic data company.  \trbcite{wu_travel-time_2004} provides a  machine-learning method based on a support vector regression (SVR) to forecast travel-times and \trbcite{quek_pop-traffic:_2006} use a fuzzy neural-network approach to address the issue of traffic data generating processes being non-linear and used fuzzy logic to improve the interoperability of their model. On the contrary,  \trbcite{rice_simple_2004}, argue that there is a linear relation between the future travel times and currently estimated conditions. They demonstrate that a regression model with time varying coefficients is capable of designing a travel time prediction scheme. Another class of forecasting models relies on classical  time series modeling. For example, \trbcite{tan_aggregation_2009} studied two classes of time series methods,  auto-regressive moving average (ARIMA) and exponential smoothing (ES). The forecasts generated by ARIMA and ES models are used as inputs to neural networks, which aggregates those into a single forecast. \trbcite{van1996combining} also proposed combing an ARIMA model with a machine learning method, the Kohonen self-organizing map, which was used as an initial classifier. \trbcite{van_lint_online_2008} addresses a parameter estimation problem of real-time learning that can improve the quality of forecasts via an extended Kalman filter for incorporating data in real-time into a parameter learning process. \trbcite{ban2011real} proposes a method for estimating queue lengths at controlled intersections, based on the travel time data measured by GPS probes. The method relies on detecting discontinuities and changes of slopes in travel time data. Another data-mining based approach for queue state estimation was proposed in \trbcite{ramezani2015queue}, who combined the traffic flow shockwave analysis with data mining techniques.

\trbcite{cheng2012exploratory}  proposes a threshold-based critical point extraction algorithm, with a goal to reduce the communication cost in the future real-time probe data collection application. A probabilistic approach, proposed in \trbcite{hofleitner2012learning} uses a learning algorithm to infer the density of vehicles on arterial roads. A heuristic filter that allows combining data from multiple types of sensors with different spatial and temporal resolutions was proposed in \trbcite{van2010robust}.

Our main contribution  is an algorithm for quick detection of the break-downs and recoveries in the traffic flow regimes. In many cases, we can  detect the breakdown  new measurement arrives. We contrast the performance of our filter with some other approaches in the ``Numerical Experiments'' subsection.


\section{Bayesian Modeling of Traffic Flow Speed}\label{sec:state-space} 
Traffic flow speed data often relies on sparse and noisy measurements. Sparseness occurs due to  a fixed grid of sensor or a dynamically changing data source such as GPS probes. Our state-space model is designed to be applicable in both scenarios. There are discontinuities in the traffic flow dynamics which need to be accounted for. We build a Dynamic Model [\trbcite{west97, carlin1992monte}]  for the traffic state during three regimes: free flow, breakdown and recovery. 

\begin{figure}[H]
	\begin{center}
		\includegraphics[width=0.7\linewidth]{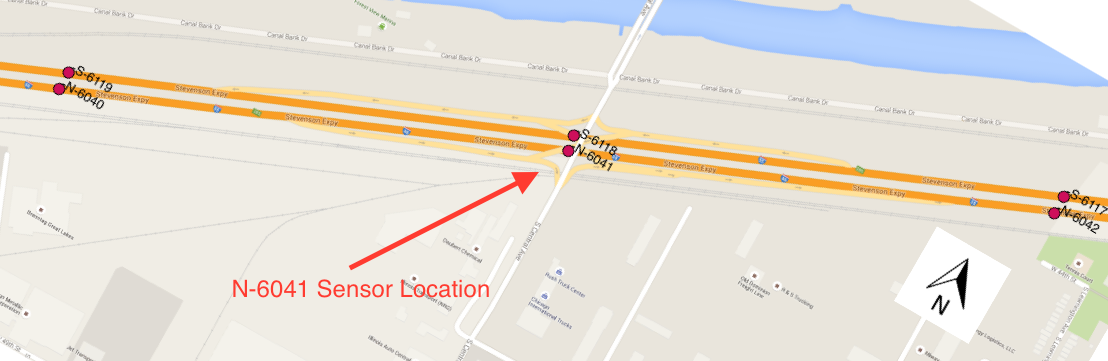}
	\end{center}
	\caption{Location of the N-6041 sensor and the geometry of the road segment.}
	\label{fig:sensor-loc}
\end{figure}

To illustrate our methodology, we use data from  a sensor on I-55 with id \textit{N-6041}. The sensor is located eight miles from the Chicago downtown on I-55 north bound (near Cicero Ave), which is part of a route used by many morning commuters to travel from southwest suburbs to the city.  As shown on Figure \ref{fig:sensor-loc}, the sensor is located 840 meters downstream of an off-ramp and 970 meters upstream from an on-ramp.

\begin{figure}[H]
\centering
\includegraphics[width=0.7\linewidth]{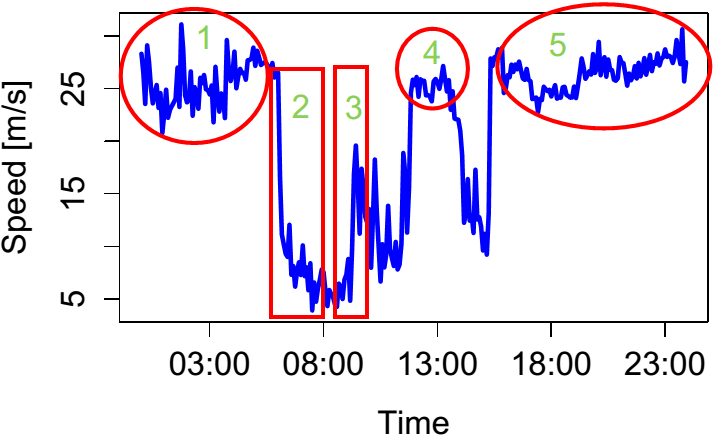}
\caption{Example of one day speed profile on May 14, 2009 (Thursday). This plot illustrates the speed profile for a segment of interstate highway I-55. Different flow regime regions were identified and labeled by the authors.}
\label{fig-day-514}
\end{figure}

Figure~\ref{fig-day-514} illustrates a typical day traffic flow pattern on Chicago's I-55 highway where sudden breakdowns are followed by a recovery to free flow regime. This traffic pattern is recurrent and   similar to one observed on other work days. We can see a breakdown in traffic flow speed during the morning peak (region 2) period followed by speed recovery (region 3). The free flow regimes (regions 1,4, and 5 on the plot) are usually of little interest to traffic managers. This data motivates our choice of the statistical model developed. The goal is to build a   model that is capable of capturing the sudden regime changes, such as free flow to congestion at the beginning of the morning rush hour (regions 1 and 2) or change in speed to the recovery regime at the end of the rush hour (regions 2 and 3).

A key modeling feature  is the inclusion of a switching state variable $\alpha_{t+1} \in \{-1,0,1\}$ to identify different flow regimes. Trends during the break-down periods and recovery periods can then be modeled using the first order polynomial component. 

Specifically, given a discretization from $t$ to $t+\Delta t$, we use a state-space model of the form
\begin{align}
	\mbox{Observation: }&y_{t+1} = Hx_{t+1}  + \gamma^Tz_{t+1} + v_{\alpha_{t+1}} , \hspace*{8ex} v_{\alpha_{t+1}} \sim N(0, V_{\alpha_{t+1}}) \label{eqn-y}\\
	\mbox{Evolution: }& x_{t+1} = 
	 G_{\alpha_{t+1}} x_t + (I-G_{\alpha_{t+1}})\mu + w_{t+1} , \  w_{t+1} \sim N(0, W_{\alpha_{t+1}})\label{eqn-state}\\
	\mbox{Switching Evolution: }&\alpha_{t+1} \sim p(\alpha_{t+1} |\alpha_{t},Z_{t})\label{eqn-alpha} 
\end{align}
where evolution gain matrix is
\[
G_{\alpha_{t+1}} = \left(\begin{array}{cc}
F_{\alpha_{t+1}} & \alpha_{t+1} \\ 
0 & 1
\end{array}  \right), ~~\mbox{and} ~\mu = \left(\begin{array}{c}v_{f}\\ 0\end{array}\right)
\]
Our hidden state variable $x_t = ( \theta_t, \ \beta_t )^T$, where $\theta_t$ is traffic flow speed and $\beta_t$ is rate of change.  We model the recovery and degradation in speed using an additive component $\alpha_t \beta_t$. We also model $\beta_t$ using a random walk model.  Measurements are then given by $y_t = (speed, \ count, \ density)_t$, and we incorporate a switching variable with three states $\alpha_t \in \{breakdown,\ free \ flow, \ recovery\}_t$. 

The observation matrix $H$, which could be time varying,  allows for partial observation of the state vector $x_t$. The parameter $v_{f}$ the  free flow speed on the road segment.
We allow for the possibility of regressors, $ z_{t+1} $, in the observation equation which effect the sensor measurement model, $\gamma$ are regressors parameters. The switching coefficient $F_{\alpha_t}$ defines weather the process is mean-reverting or not. We  define it  by 
\begin{equation*}
	F_{\alpha_t} = \left\{
	\begin{aligned}
		& 1,  \ \alpha_t \in \{1,-1\}\\
		&F_0, \  \alpha_t = 0.
	\end{aligned}
	\right.,
\end{equation*}
where $F_0 < 1$  is the rate of mean-reversion during a free flow regime. The dynamics $p(\alpha_{t+1} | \alpha_t,Z_t)$ of the switching evolution depends on the set of variables $Z_t$, that explain regime shifts. Several approaches to model switching process $p(\alpha_{t+1}|\alpha_t, Z_{t})$ and choosing $Z_t$ are described in  the next sub-section. 

The main task of our approach is to detect the start of breakdown and recovery as soon as possible.  The causes of breakdown or recovery might be different. For example a breakdown might  be caused  by a recurrent demand that increases capacity or non-recurrent conditions, such as weather or traffic accidents. Characteristic change in the traffic flow speed would be the same in either situation and thus our model is agnostic to the cause of the start or end of a congestion period. 
		
\subsection{Modeling $p(\alpha_{t+1}| \alpha_t, Z_t)$}\label{sec-alpha}
The discrete state $ \alpha_{t+1} \in \{ -1,0,1 \} $ models breakdown ($\alpha = -1$), free flow ($\alpha = 0$), and recovery ($\alpha = 1$) regimes of a traffic system. We need to specify a transition kernel for the evolution of this state given a set of exogenous predictor variables, denoted by $Z_t$, and current state $\alpha_t$.  Given $Z_t$, the  set of probabilities,  $p(\alpha_{t+1}| \alpha_t, Z_t)$,  will then form a $3 \times 3$ matrix which will be combined with the evolution of the hidden state vector $x_t$. 

\begin{figure}[H]
	\includegraphics[width = 1\linewidth]{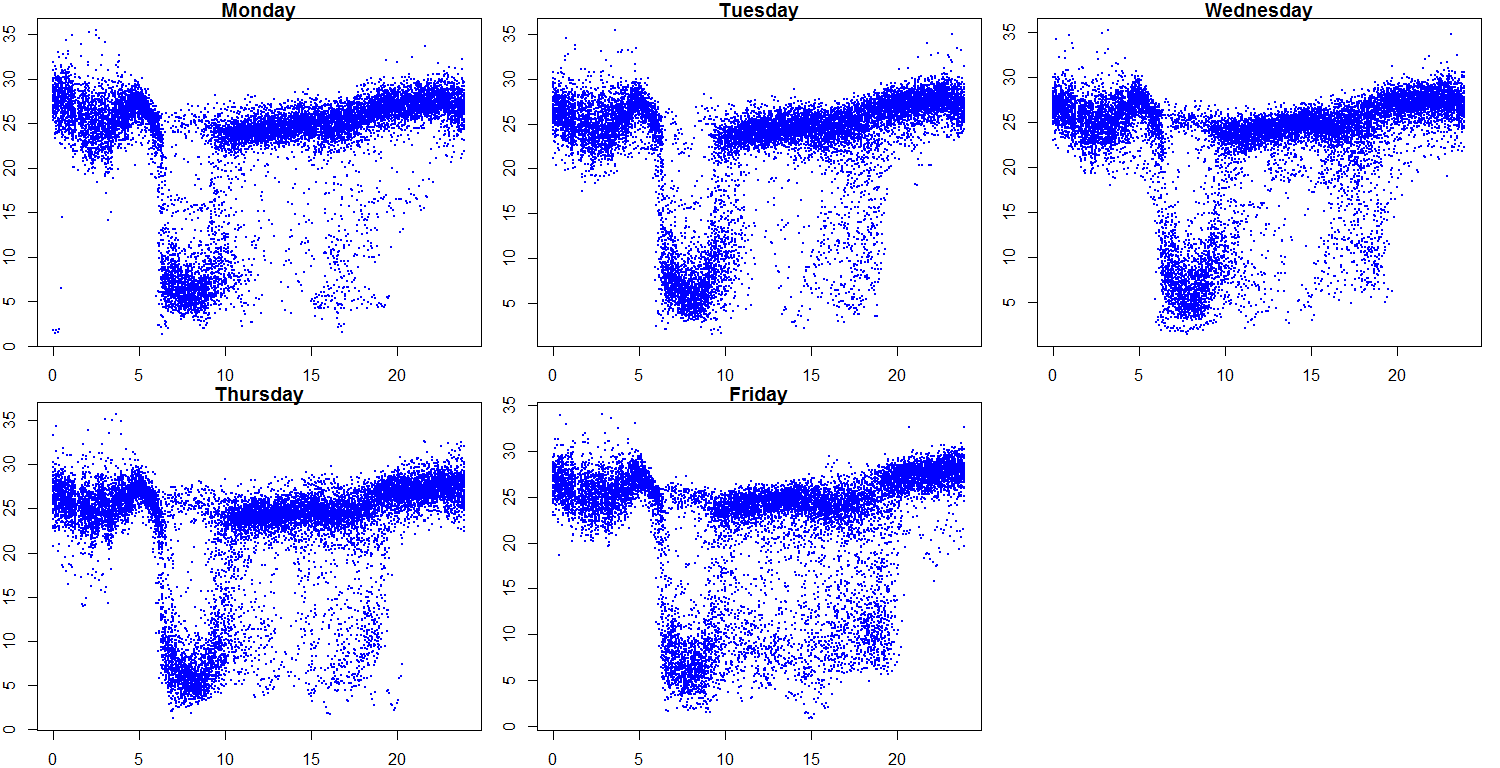}
	\caption{Daily Traffic Patterns, measured in 2009. Each plot shows raw speed measurements averaged over five minute intervals from the sensor N-6041 for a given day of the week, with holidays and days with erroneous  measurements removed from data.}\label{fig:daily-traffic}
\end{figure}

Figure \ref{fig:daily-traffic} shows that break downs  happen more or less the same time during morning or evening peak periods on a work day. Therefore, it is  natural to introduce an exogenous variable $Z_t =  (period, day \ of \ week), \ period \in \{1,2,3\}$ where the three periods correspond to morning, evening peak period and the rest of the day. Incorporating additional considerations beyond a period of the day and day of the week, such as weather, month, and special event leads to 
\[
Z_t = (day \ of \ week, month,  weather, month, event, accident)_t.
\]
One way to build such a model is to identify   $3 \times 3$ transition matrix $p(\alpha_{t+1}| \alpha_t, Z_t)$ for each combination of the parameters in $Z_t$ based on the historic observations.

\begin{figure}[H]
	\begin{tabular}{cc}
	\includegraphics[width=0.5\linewidth]{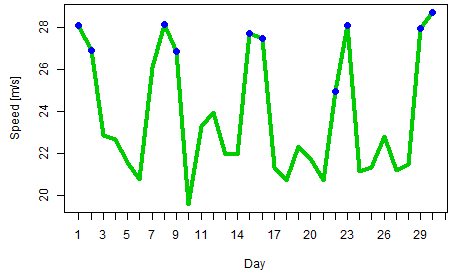} & \includegraphics[width=0.5\linewidth]{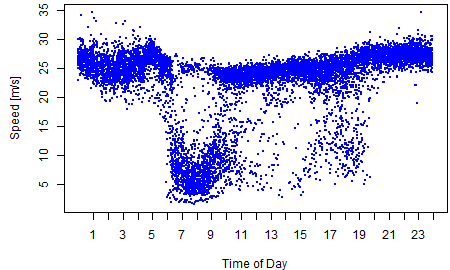}\\
	(a) Average Speed per day (April, 2009) &(b) Speed on Wednesdays (2009)
\end{tabular}
\caption{Recurrent traffic flows. The left panel (a) shows average speed as measured by the sensor N-6041 for each day during the April 4 - March 3 of 2009 time period. Weekends are marked by blue dot. The right panel (b) shows raw speed measurements from the sensor N-6041 for each five minute interval of every Wednesday in 2009.}
\label{fig:avg-speed-n6041-2009}
\end{figure}

Figure~\ref{fig:avg-speed-n6041-2009}(a) shows the average speed for a fixed location on the network for the April 4 - March 3 period in 2009, weekend days identified by blue dots. We see that the average speed on weekend is very close to free flow speed, roughly 63 mi/h, which means there is no congestion on those days. On the other hand average speed on a work day is significantly lower as a result if congestion during rush hour. We  also can see an unusual average speed on April 10 of 2009 (seventh observation). It was  Good Friday. Thus, far less traffic was observed compared to a typical Friday.

Figure~\ref{fig:avg-speed-n6041-2009}(b) shows a scatter plot of measured traffic flow speed for all non-holiday Wednesdays in 2009. The measurements are taken every five minutes. We can see that congestion starts roughly at the same time at around six in the morning and lasts roughly for three hours. The  breakdown in the evening happens less frequently. 

Another effect to be modeled is  traffic congestion that is a recurrent event, and is similar from one week to another. Figure~\ref{fig:daily-traffic} shows the recurrent traffic conditions grouped by the day of the week. This observation can be used to choose $Z_t$. In particular, an approach based on  non-parametric regression that uses historical traffic flow data. For example,  \trbcite{smith02} and \trbcite{chiou2013}  showed that the last three measured speed values, perform very well for predicting traffic flows. In this case we can write $Z_t$ as
\[
Z_t = (\alpha_t, \hat{x}_t, \hat{x}_{t-1}, \hat{x}_{t-2}, \ time \ of \ day)^T,
\]
where $\hat{x}_t$ is the filtered value of state vector $x_t$.
Then the transition dynamics $p(\alpha_{t+1} |\alpha_t, Z_t)$ is generated by computing a weighted average of those points from historic database that fall within a neighborhood of $Z_t$. To calculate the neighborhood, we calculate distance between $Z_t$ and each of the point from the data base, and then choose $k$ points with smallest distance. The weights are proportional to the distance.

\section{Tracking Traffic Flows during Special and Weather Events}\label{sec:tracking}
\begin{figure}[H]
	\begin{tabular}{cc}
		\includegraphics[width=0.5\linewidth]{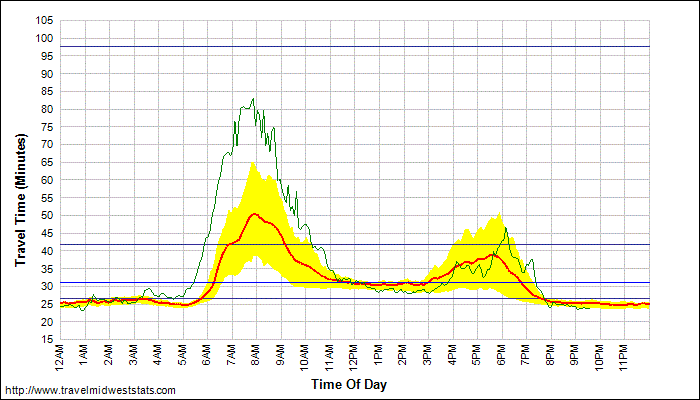} & \includegraphics[width=0.5\linewidth]{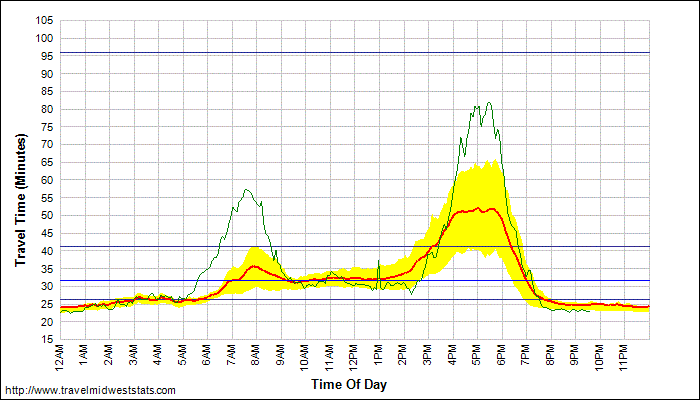}\\
		(a) North Bound & (b) South Bound
	\end{tabular}
	\caption{Impact of light snow on travel times on I-55 near Chicago on December 11, 2013. Both plots show  travel time on a 27 mile stretch of highway I-55 between I-355 and I-94 in both north and south directions. The north bound direction is from southwest suburbs to the city. The red line is a travel time averaged over previous 150 days, the yellow area show 70\% confidence interval for the data and green line is the travel time on the day of the event. Source: \url{www.travelmidweststats.com}}\label{fig:light-snow}
\end{figure}

Our model is flexible in the sense that we can handle special  events or severe weather conditions that can upset a typical traffic patterns.  To show the empirical effect of snowy weather, Figure~\ref{fig:light-snow} compares the expected travel time (red line), which is calculated based on historical data for the last 150 days, with the travel time on a snow day (green line), for December 11, 2013. There were 1.8 inches of snow on this day, with snow starting at midnight and continuing till noon. There were no traffic accidents on this road segment on this day. As we can see, even a light snow in a region, where drivers are used to driving during snow days can cause major delays. The yellow region is the 70\% confidence interval based on historical data.

\begin{figure}[H]
	\begin{tabular}{cc}
		\includegraphics[width=0.5\textwidth]{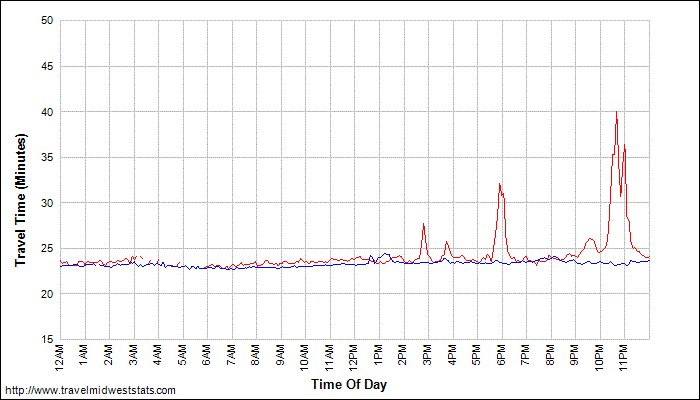} & \includegraphics[width=0.5\textwidth]{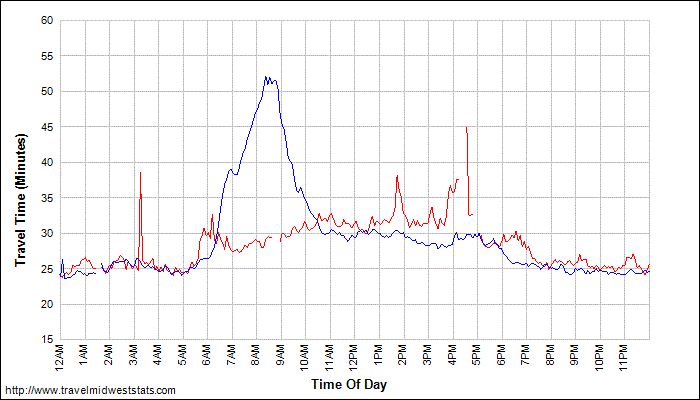}\\
		(a) NATO Summit on Sunday May 20, 2012 & (b) NATO Summit on Monday May 21, 2012 \\
		\includegraphics[width=0.5\textwidth]{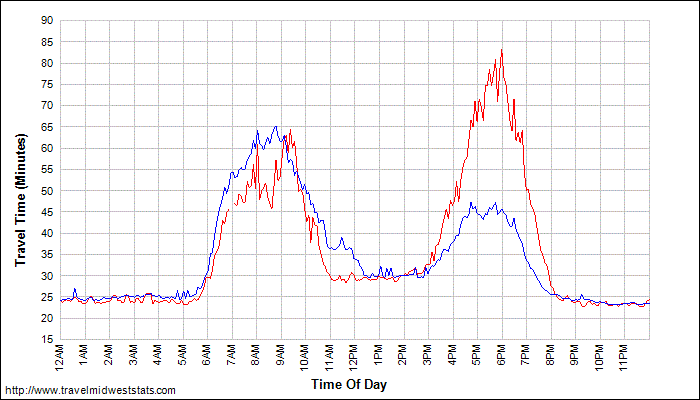} & \includegraphics[width=0.5\textwidth]{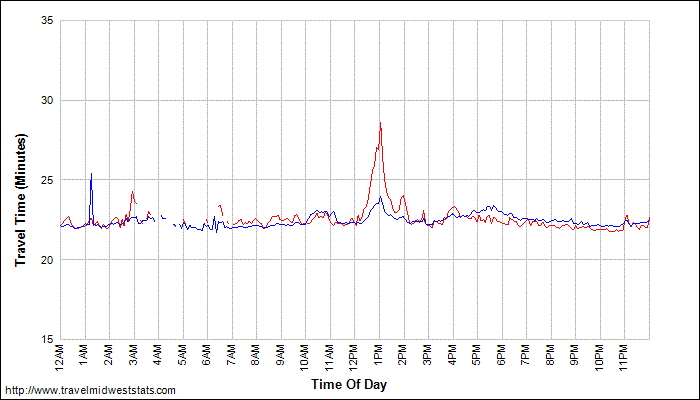}\\
		(c) New York Giants at Chicago Bears on  & (d) Baltimore Ravens at Chicago Bears on \\
		Thursday October 10, 2013 & Sunday November 10, 2013
	\end{tabular}
	\caption{Impact of special events on I-55 north  bound travel times.  All plots show  travel time on a 27 mile stretch of highway I-55 between I-355 and I-94 in both north  (towards the city) and south directions. All plots compare travel times on the day of the special event (red line) with the travel time on the same day of the week averaged over the previous 150 days (blue line). \url{www.travelmidweststats.com}}\label{fig:special-events}
\end{figure}

Special events is another potential cause of unusual traffic conditions. Figure~\ref{fig:special-events} shows impacts of special events on travel times on interstate I-55 north bound (towards the city).  The weekday football game, which takes place at Solder Field stadium in Chicago downtown, combined with typical commute traffic has  a very significant impact on travel times. Weekend special events have a relatively minor negative impact. On the other hand, the NATO summit that was held in Chicago's McCormick Place located slightly south of downtown on Monday, had positive impact on travel times. This can be explained away by regular commuters, who knew about the event, changing their departure times, using  commuter rail or simply working from home on this day.

\section{Building a Particle Filter}\label{sec:pf}
Given our state-space  model, the goal is to provide an on-line algorithm, for finding, in an on-line fashion, the set of joint filtered posterior distributions $p(x_t, \alpha_t | y_{1:t})$ using a two-step procedure  for both $x_t, \alpha_t$, the traffic flow state, and switching variable at each time point. The advantage of particle filtering is that we simply re-sample from the predictive distribution of the current posterior and then  propagate, to the next set of particles, to generate the approximation to the conditional posterior update. Appendix~\ref{app:pf} provides a review of particle filtering and learning methods.

From a probabilistic viewpoint, we can re-write equations  (\ref{eqn-y}) - (\ref{eqn-state}) as a hierarchical model
\begin{align}
	 (y_{t+1}|x_{t+1})&\sim N(Hx_{t+1} + \gamma^T z_{t+1},V_{\alpha_{t+1}})\label{measurement-probability}\\
	(x_{t+1} | x_t, \alpha_{t+1})& \sim N(G_{\alpha_{t+1}}x_t + (I-G_{\alpha_{t+1}})\mu,  \  W_{\alpha_{t+1}}). \label{evolution-probability}
\end{align}

Now, suppose that we are currently at time $t$. We assume a particle approximation for the joint posterior of $(x_t, \alpha_t)$ is of the form
\[
p^N (x_t , \alpha_t | y^t) = \sum_{i=1}^N w^{(i)}_t \phi(x_t, m_t^{(i)}, C_t^{(i)})  \delta_{\alpha_t^{(i)}} (\alpha_t) 
\]
where $y^t = (y_1,...,y_t)$ and $ \delta_\alpha(\cdot)$ is a Dirac measure and $  w^{(i)}_t $ is a set of particle weights, which we will provide recursive updates for. We denote the Kalman moments by $s_t=(m_t, C_t)$ which form a set of conditional sufficient statistics for the state. We describe the recursions later. Here $ \phi( x , \mu , C) $ denotes a normal density with mean $\mu$ and covariance $C$
\[
\phi(x, \mu,C) = \frac{1}{ \left(2\pi\right)^{\frac{k}{2}} |C|^{\frac{1}{2}}}\exp \left(-\frac{1}{2}(x-\mu)^TC^{-1}(x-\mu)\right).
\]

To assimilate the next measurement, we need to find an  updated  posterior for $(x_{t+1}, \alpha_{t+1})$, with approximate weights $w_{t+1}^{(i)}$ and particles $(x_{t+1}, \alpha_{t+1})$ of the form
\[
p^N \left (x_{t+1}, \alpha_{t+1}|y^{t+1} \right ) = \sum_{i=1}^N w_{t+1}^{(i)} \phi\left (x_{t+1},  m_{t+1}^{(i)}, C_{t+1}^{(i)} \right ) \delta_{\alpha_{t+1}^{(i)}} (\alpha_{t+1}),
\]
Our weights will be updated using the predictive likelihood, $ p( y_{t+1} | s_t^{(i)} ) $, which are re-normalized. 
We aim to provide a posterior with the same number of particles $N$ with mixture weights of the form 
\[
w_{t+1}^{(i)}  = \frac{ w^{(i)}_t  p( y_{t+1} | ( \alpha_t , s_t )^{(i)} ) }{\sum_{i=1}^N w^{(i)}_t   p( y_{t+1} | ( \alpha_t , s_t )^{(i)} ) } \; .
\]

\subsection{Recursive Updating}\label{subsection-traffictracking}
At time zero, we set an initial state distribution $p(x_0 | \alpha_0)$ as follows
\[
p(x_0 | \alpha_0) = \sum_{i=1}^{N} w_{\alpha_0}^{(i)} \phi(x_0,\mu_{\alpha_0}^{(i)}, C_{\alpha_0}^{(i)})
\]
We take $ p(\alpha_0 = s)  = 1/3,~s \in \{-1,0,1\}$.

Conditional on the hidden switching state $\alpha_t$, the Kalman filter recursions, imply that filtered posterior distribution at time $t$ is also mixture multivariate normal, i.e 
$$ p^N(x_t |\alpha_{1:t}, y_{1:t})=\sum_{i=1}^{N} w^{(i)}_t \phi(x_t,m^{(i)}_t, C^{(i)}_t) $$ where $ (m_t,C_t)^{(i)} $ are functions
of the whole path $ \alpha_{1:t},y_{1:t} $. 

The goal is to find the next filtered posterior $p(x_{t+1} | y_{1:t+1} )$, which is  obtained from the marginal of the joint posterior $ p(x_{t+1} , \alpha_{t+1}| y_{1:t+1} )$.

Given, that $x_t \sim N(m_t, C_t)$, from evolution equation (\ref{evolution-probability}), it follows that 
\begin{align*}
p(x_{t+1}|\alpha_{t+1}, s_t) = &\phi(x_{t+1},G_{\alpha_{t+1}}m_t+(I -G_{\alpha_{t+1}})\mu, G_{\alpha_{t+1}}C_tG_{\alpha_{t+1}}^T +  W_{\alpha_{t+1}})
\end{align*}

To implement this algorithm, first compute the predictive likelihood of the next observation $y_{t+1}$ given $\alpha_{t+1}, s_t$, where $s_t = (m_t, C_t)$ is the sufficient statistics. This can be done by marginalizing out $x_{t+1} $ from measurement equation (\ref{measurement-probability}). We obtain a marginal predictive distribution
\begin{align*}
p(y_{t+1}|\alpha_{t+1},s_t)   = &\int p(y_{t+1}|\alpha_{t+1},x_{t+1})p(x_{t+1}| \alpha_{t+1}, s_t)dx_{t+1} \\
= & \int \phi(y_{t+1} , Hx_{t+1}+ \gamma^T z_{t+1},V_{\alpha_{t+1}})  \\
&  \phi(x_{t+1} , G_{\alpha_{t+1}}m_t+ (I - G_{\alpha_{t+1}})\mu, G_{\alpha_{t+1}}C_tG_{\alpha_{t+1}}^T + W_{\alpha_{t+1}})dx_{t+1} \\ 
= &\phi\left(y_{t+1} ,  \mu^y_{t+1} , V^p_{t+1}\right).
\end{align*}
where $\mu^y_{t+1} = H(G_{\alpha_{t+1}}m_t+ (I - G_{\alpha_{t+1}})\mu) + \gamma^T z_{t+1}$, $V^P_{t+1} = V_{\alpha_{t+1}}+H^T\left(G_{\alpha_{t+1}}C_tG_{\alpha_{t+1}}^T +  W_{\alpha_{t+1}}\right)H$
and $V_{\alpha_{t+1}}$ and $W_{\alpha_{t+1}}$ are given variance-covariance matrices.

We can further marginalize out $ \alpha_{t+1} $ using the transition kernel $p( \alpha_{t+1} | \alpha_t , Z_t) $. This leads to  3-component mixture predictive  model of the form
$$
p(y_{t+1}| x_t , \alpha_{t})  = \sum_{\alpha_{t+1} \in \{0,1,-1\}} 
\phi ( y_{t+1}, \mu^y_{t+1} ,  V^p_{t+1} ) p(\alpha_{t+1}|\alpha_t,Z_t) 
$$
Our model, therefore allows for heavy-tails and non-Gaussianity in the traffic flow evolution. We now show how this can be used to implement a particle filter and learning algorithm and track the filtered posterior distributions of the hidden state $p(x_t | y_{1:t} )$ over time as new data $y_{t+1} $ arrives.

Given $ (\alpha_{t+1}, y_{t+1} )$, we need to update $ s_{t} = (\mu_{t},   C_{t})^T $, where we suppress the index $i$ for clarity.  

These updates are given by Kalman recursion operator, $ \mathcal{K} $, which is given by
\begin{equation*}
\begin{tabular}{ll}
$\mu_{t}^f = G_{\alpha_{t+1}}\mu_t+(I-G_{\alpha_{t+1}})\mu,$ & $C_{t}^f =  G_{\alpha_{t+1}}C G_{\alpha_{t+1}}^T +  W_{\alpha_{t+1}}$  \\ 
$\mu_{t} = \mu_{t}^f + K_t(y_t - H\mu_t^f ),$ & $C_t = (I - K_tH_t)C_t^f$
\end{tabular} 
\end{equation*}
with Kalman gain matrix $$K_t = C_t^fH^T(HC_t^f H^T + V)^{-1} $$. 

Now we are in a position to find the predictive density in equation (\ref{eqn-predictiveweight}), namely $p(y_{t+1}|\alpha_t, s_t) $, which  is a 3-component mixture of Gaussians. This will lead to an efficient Rao-Blackwellised particle filter \\

\textbf{Algorithm. Particle Filtering for traffic flows}:

\textit{Step 1} (Draw) an index $k_t(i) \sim  Multi ( w_{t+1} )$-distribution with weights
\begin{equation}\label{eqn-predictiveweight}
	w_{t+1}^{(i)}  =  p\left(y_{t+1}| s_t^{(i)}\right )/\sum_{i=1}^{N}p\left(y_{t+1} | s_t^{(i)}\right) .
\end{equation}

\textit{Step 2} (Propagate) switching state $ \alpha_{t+1}^{(i)} \sim p( \alpha_{t+1} | \alpha_t^{k_t(i)} ) $ 

\textit{Step 3} (Propagate) sufficient statistics  $s^{k_t(i)}_t$ using assimilated data and Kalman filter recursion
\begin{equation}\label{eqn-kalmanrecursion}
	s_{t+1}^{(i)} = \mathcal{K}(s_t^{k_t(i)},\alpha_{t+1}^{(i)}, y_{t+1})
\end{equation}

The weights are updated according to the following rule
\[
w_{t+1}^{(i)}  = \frac{ w^{(i)}_t  p( \alpha_{t+1} | \alpha_t^{(i)} ) }{\sum_{i=1}^N w^{(i)}_t p( \alpha_{t+1} | \alpha_t^{(i)})  }
\]
Finally, we draw new state vector $ x_{t+1} $ from its mixture multivariate normal distribution.

\section{Tracking Traffic Flow on Interstate I-55}\label{sec:example}
\subsection{Dataset Description}
The data was provided by the Lake Michigan Interstate Gateway Alliance\\ (\textit{http://www.travelmidwest.com/}) formally Gary-Chicago-Milwaukee Corridor (GCM). The  data are measurements from the loop-detector sensors installed on interstate highways. A loop detector is a very simple presence sensor that senses when a vehicle is on top of it and generates an on/off signal. There are slightly more then 900 loop-detector sensors that cover a large portion of the Chicago metropolitan area. Every 5 minutes a report for each of the loop detector sensors is recorded.  Our data contains averaged \textit{speed}, \textit{flow}, and \textit{occupancy}.  Flow is defined as the number of off-on switches. Occupancy is defined as percentage of time a point on a road segment was occupied by a vehicle, thus it varies between 0 (empty road) to 100 (complete stand still).  We assume a constant vehicle length, and treat speed derived from a single loop-detector occupancy measurement as the true traffic flow speed.

\subsection{Numerical Experiments}
Consider a single road segment. We  use measured data for a 24-hour period taken on a week day. The segment we consider is a part of highway I-55 north bound. This part of the highway is heavily congested during the morning rush hour, mostly due to commuters, who travel from the south-west suburbs to the central business district of Chicago. There were no special events on that day and the weather was clear, thus a very similar congestion pattern can be observed on any ``usual'' work day on this road segment. The measurements are made by a single loop detector. For this example we calculate the filtering distribution for the travel speed, flow regime, and rate of change variables on this segment. The time series of measured speeds is of length $N= 288  \ (24\times 12)$.

The state $x_t = (\theta_t, \ \beta_t)^T \in \mathbb{R}^2$ is a true travel speed and associated trend coefficient. The parameters of the observation and evolution models are set to:
\[
\ H = (1 \ 0), \  V_{\alpha_t} = 4, \ F_0 = 0.5 , \ W_{\alpha_t} = \left(\begin{array}{cc}
1.9 & 0 \\ 
0 & 4.5
\end{array} \right),
\] 
Given that we did not have access to manufacturer's specifications of the loop detectors,  we use a value for the measurement error  within the guidelines of the specification.  The error for the evolution equation was chosen using maximum a posteriori mode estimation based on the data from the previous 30 days.

In our simulation study, we define a transition matrix $P$ with equilibrium probabilities $\pi P= \pi$ that calibrate well to the three states in practice. Thus, we construct a Markov-switching process with the following probability transition matrix
\[
P_{\alpha_t} = \left(
\begin{tabular}{@{\extracolsep{5pt}} ccc} 
$0.6$ & $0.3$ & $0.1$ \\ 
$0.15$ & $0.7$ & $0.15$ \\ 
$0.3$ & $0.1$ & $0.6$ \\ 
\end{tabular}  
\right),
\]
where $\alpha_t \in \{breakdown,\ free \ flow, \ recovery\}$. The transition probabilities were fitted using a maximum a posteriori mode estimator. $F_0$ was fitted using data from times when traffic is stationary, and $W_{\alpha_t}$ was fitted using the data from both the stationary and non-stationary regimes.

\begin{figure}[H]
\centering
\includegraphics[width=0.9\linewidth]{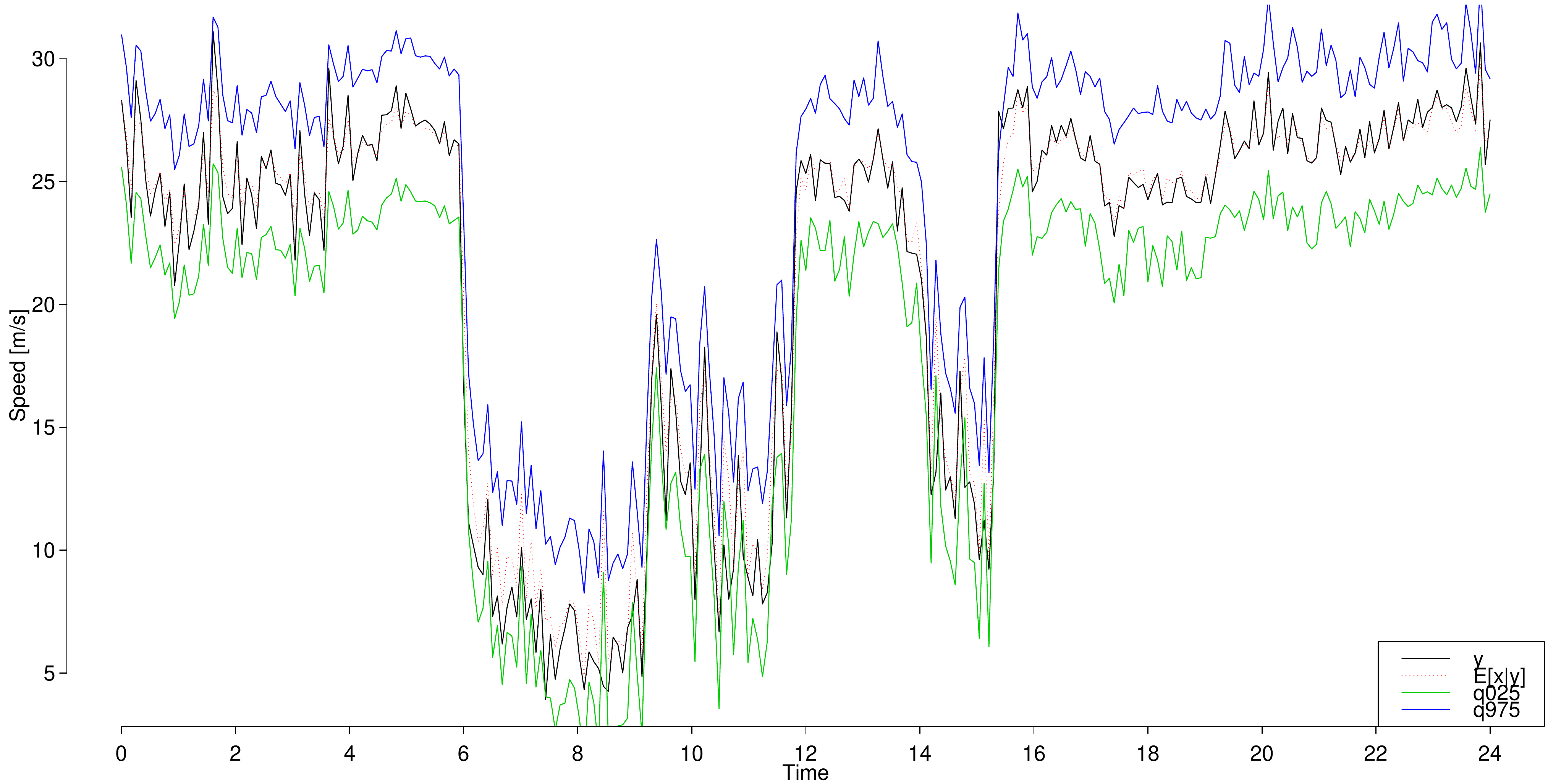}
\caption{Filtered traffic speed given loop detector measurements.}
\label{fig:filtered_speed}
\end{figure}

Figure~\ref{fig:filtered_speed} shows the filtered speed and its quantiles, along with measured data. We see that the filtered state curve more-or-less follows the measurement curve. The evolution model proposed in this paper is very general and allows large changes in the speed state. Thus, the jittering behavior of measurement get mostly explained by statistical model and does not get filtered out. 

\begin{figure}[H]
\centering
\includegraphics[width=0.9\linewidth]{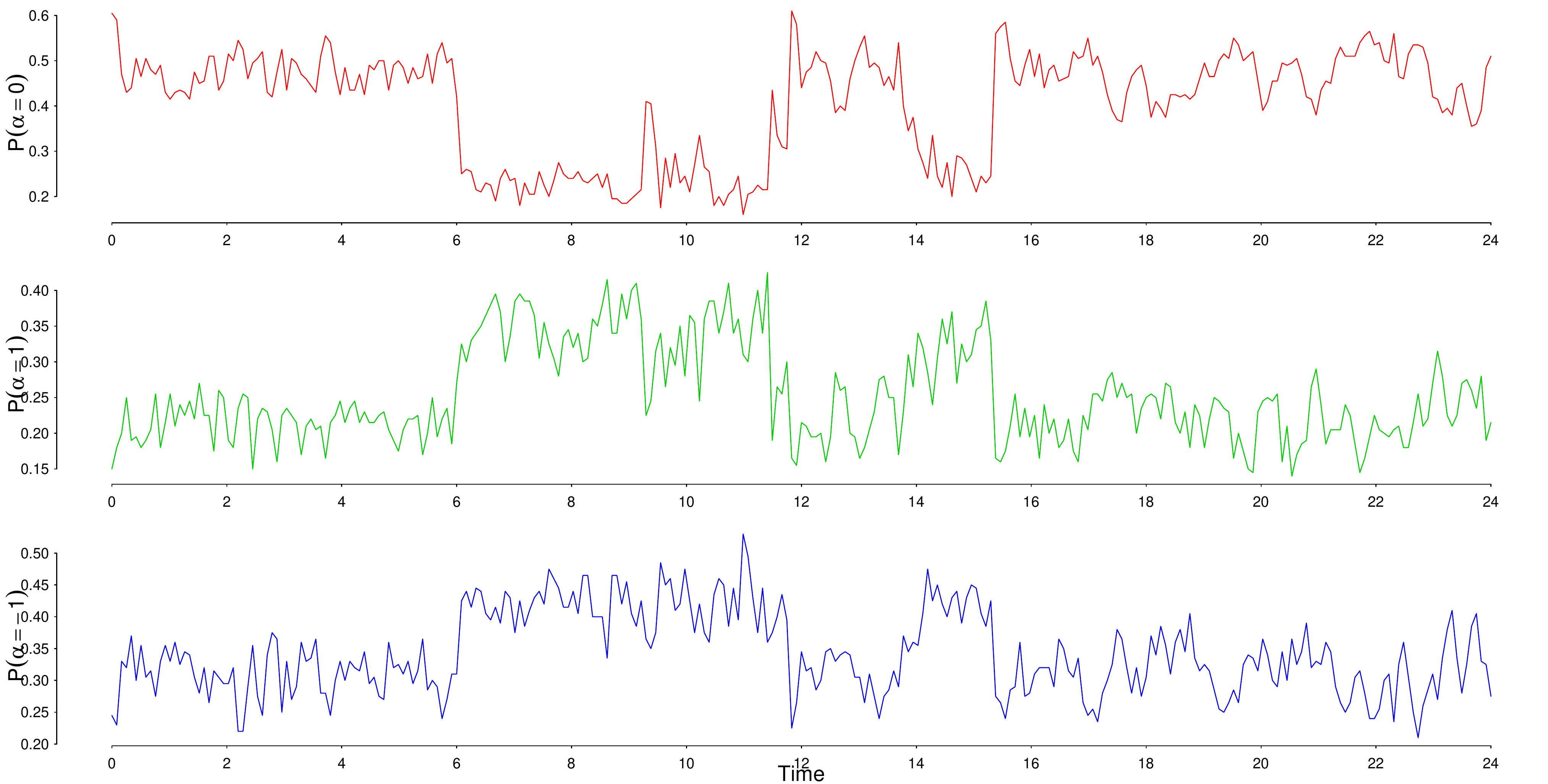}
\caption{Filtered value of $P(\alpha = i), ~ i=1,2,3$}
\label{fig:filtered_alph}
\end{figure}

Figure~\ref{fig:filtered_alph} shows the filtered probability  of $\alpha$ for each of the values (0,1,-1). We can see that the algorithm accurately captures the changes in the flow regime,  assigns high probability to free flow regime before the morning peak and the shifts probability to the breakdown regime.  

\begin{figure}[H]
\centering
\begin{tabular}{cc}
	\includegraphics[width=0.5\linewidth]{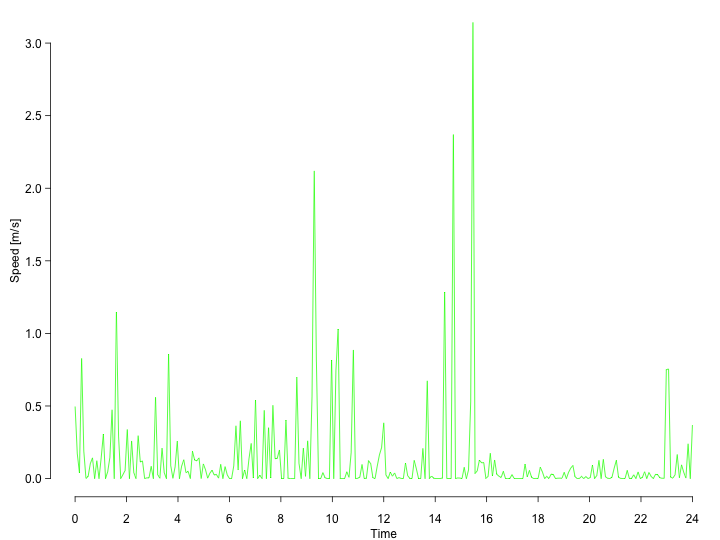}&  \includegraphics[width=0.5\linewidth]{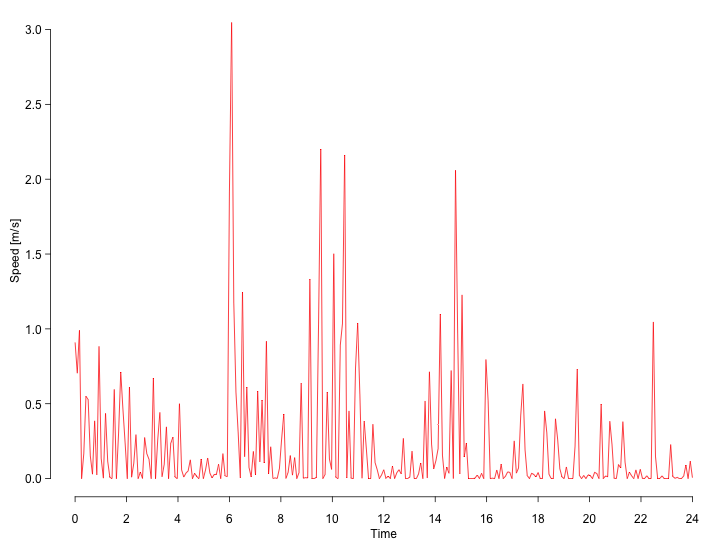}\\
	(a) Recovery & (b) Degradation \\
\end{tabular} 
\caption{Rate of change  of traffic flow}
\label{fig:beta}
\end{figure}

Figure~\ref{fig:beta} shows the filtered values for rate of change of speed during recovery and breakdown regimes. The algorithm captures all of the changes in traffic flow change rates. The algorithm captures ``fast'' breakdown a little after 6am and ``slow'' breakdown at around 10am. It also captures, for example, the recovery between 2pm and 4pm.

\begin{figure}[H]
	\centering
	\begin{tabular}{cc}
		\includegraphics[width=0.5\linewidth]{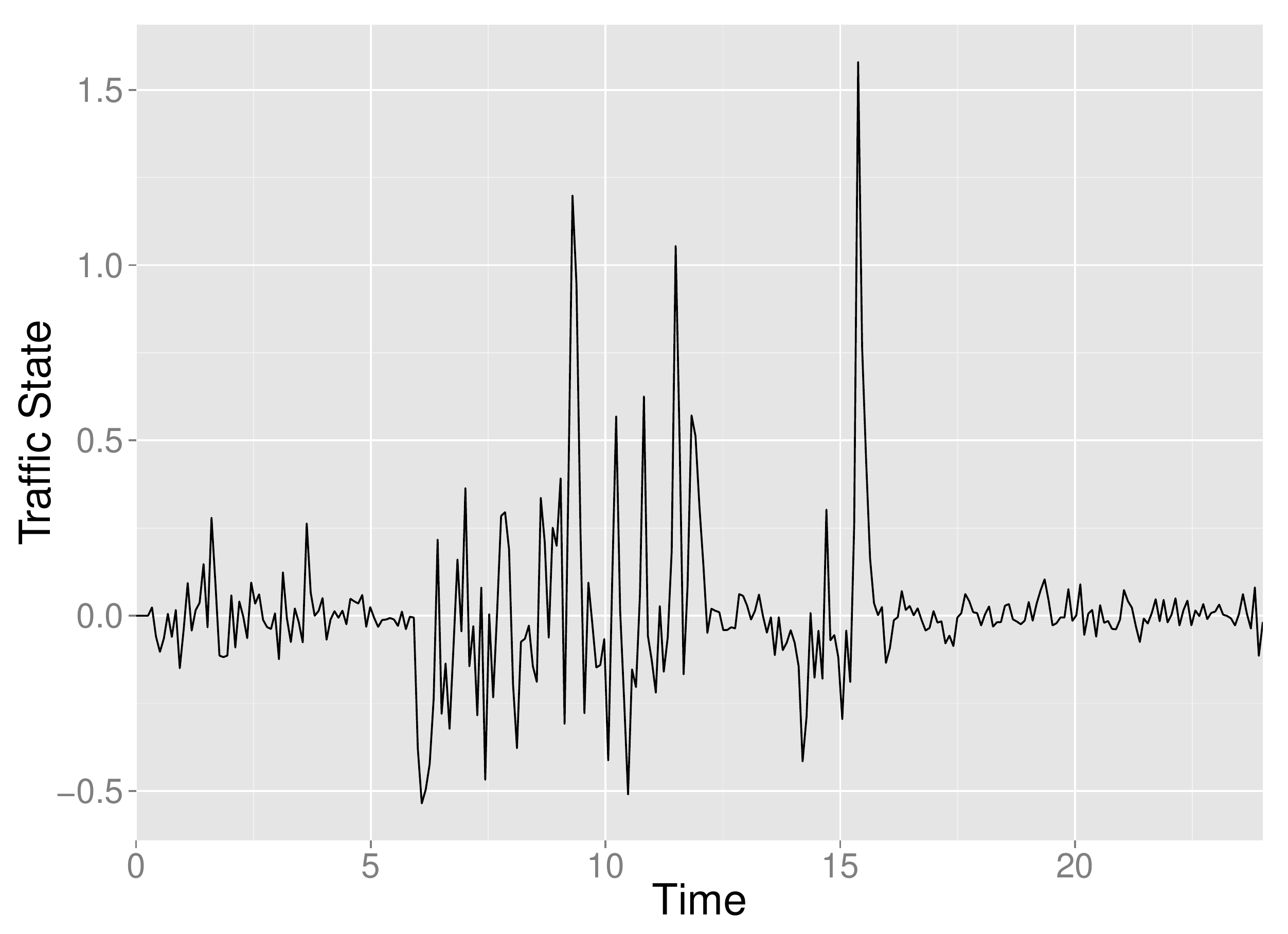}&  \includegraphics[width=0.5\linewidth]{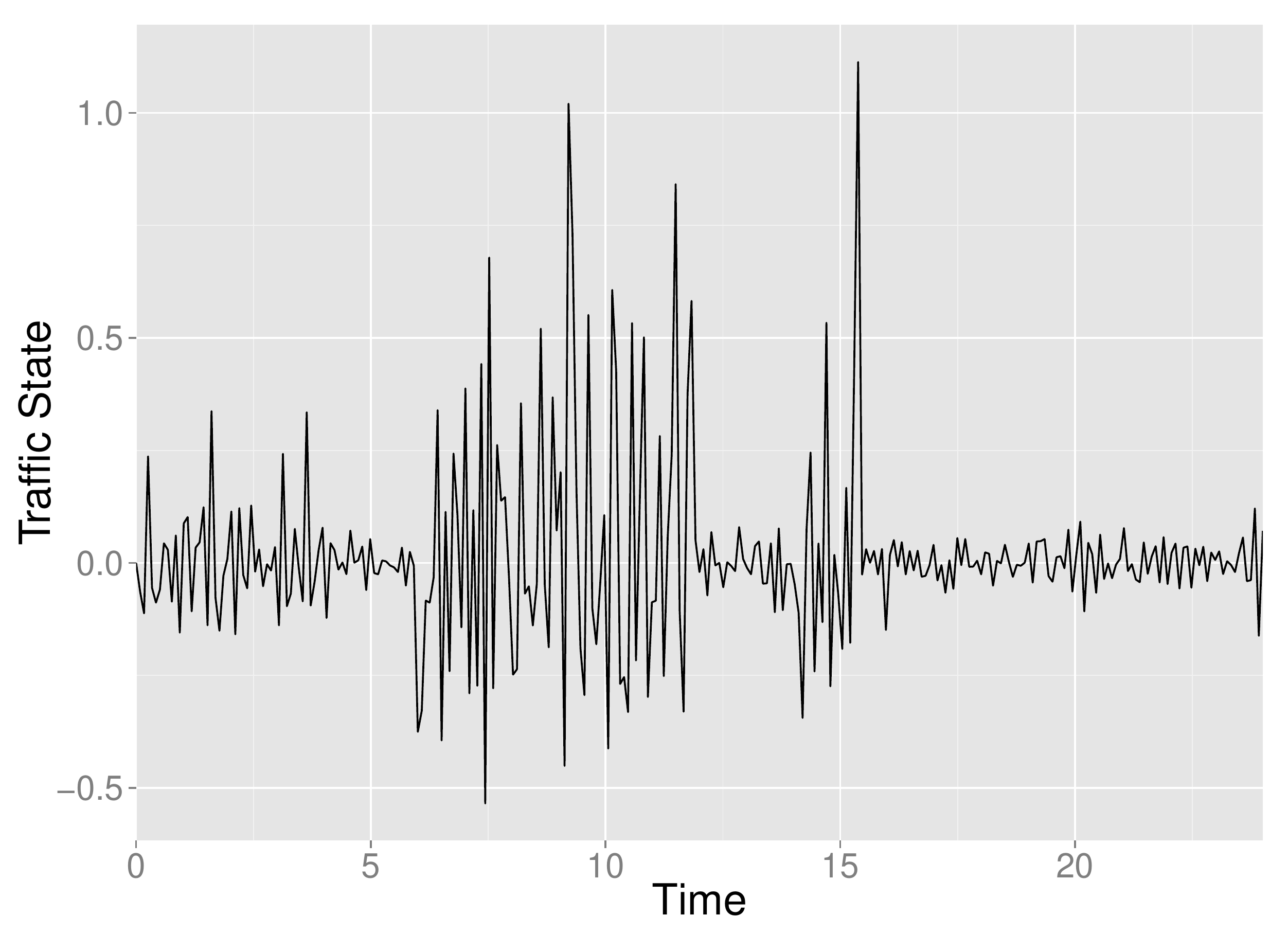}\\
				(a) Naive Filter 1 & (d) Naive Filter 2 \\
				\includegraphics[width=0.5\linewidth]{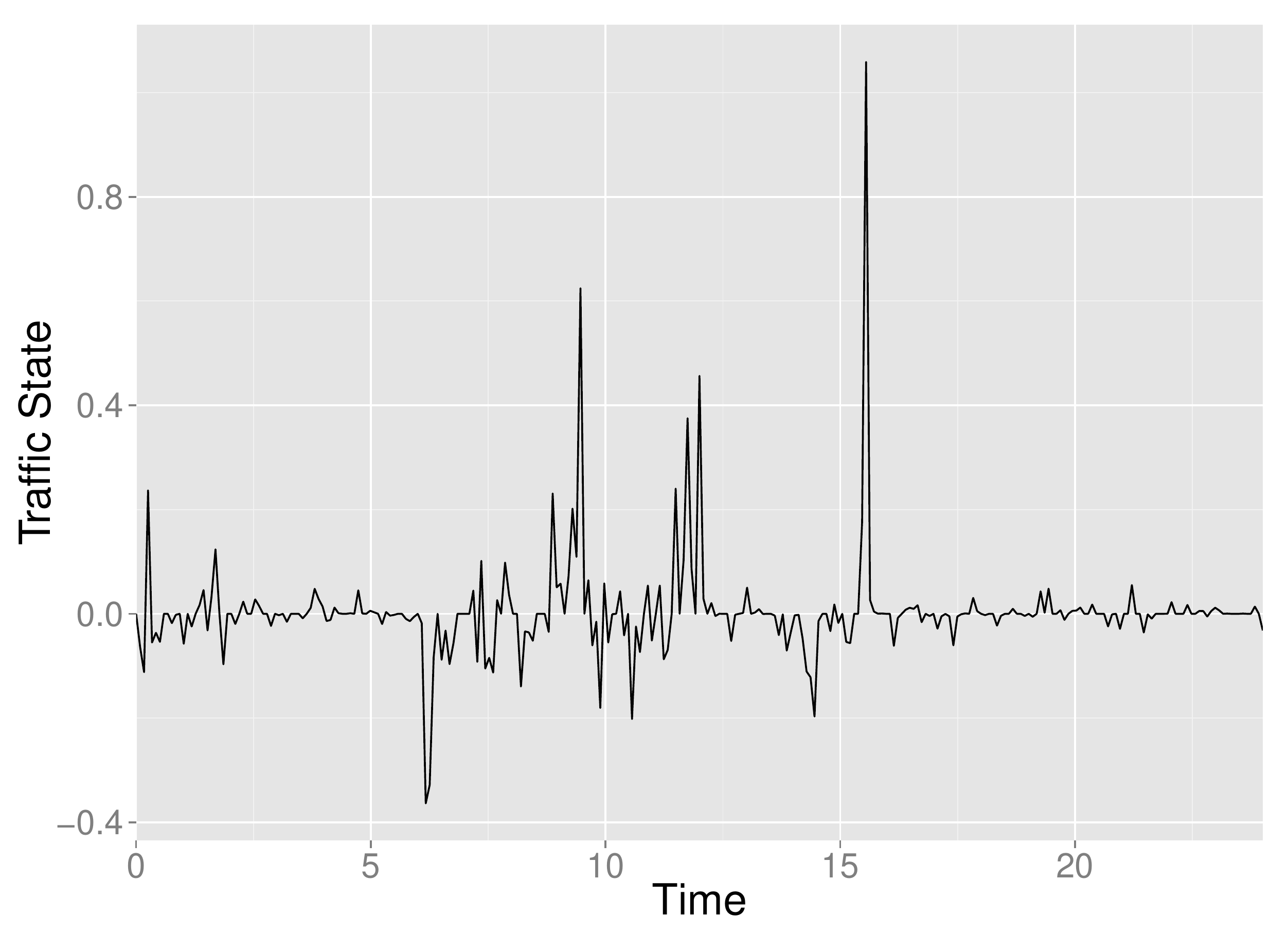}&  \includegraphics[width=0.5\linewidth]{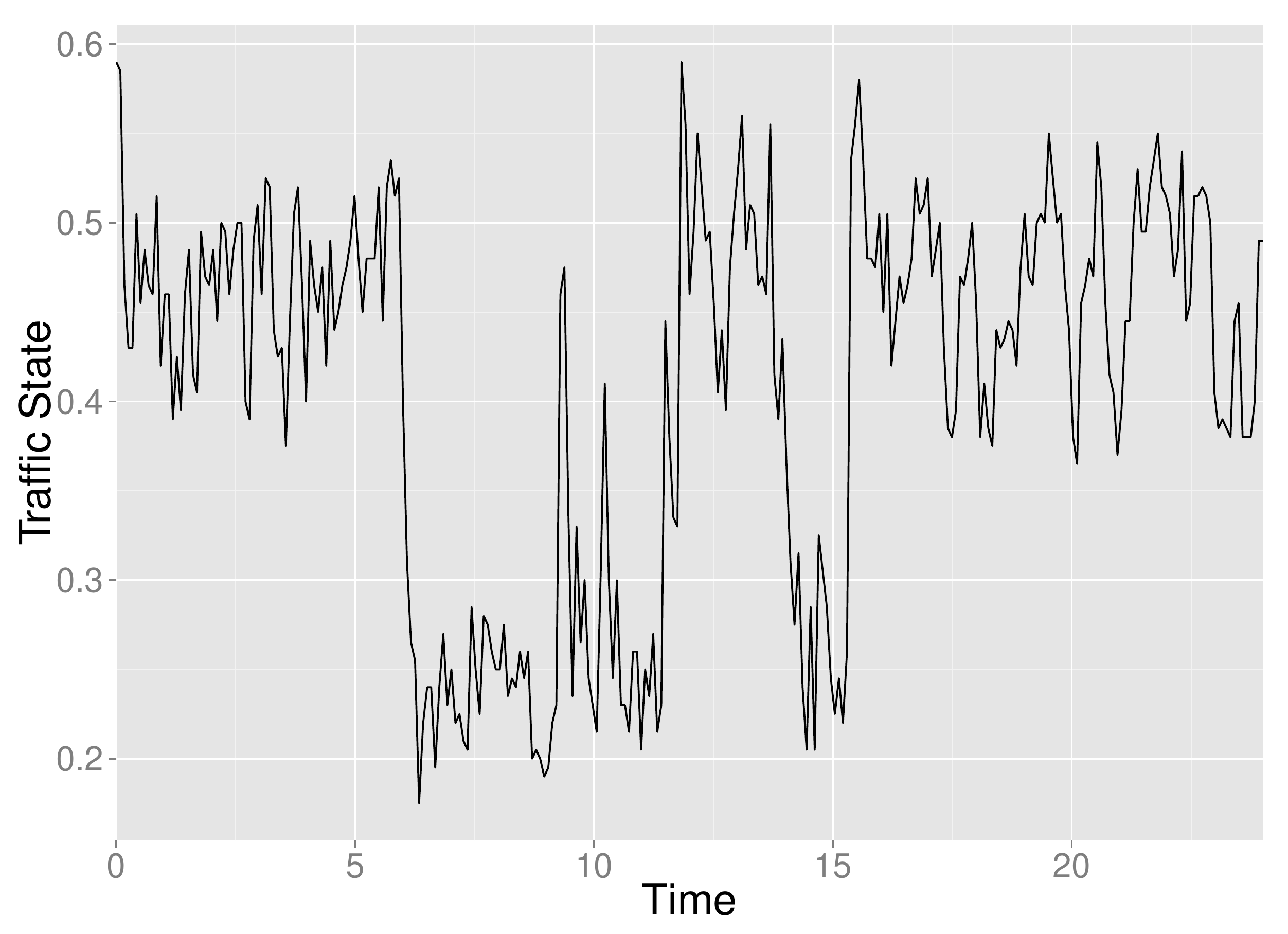}\\
		(c) Naive Filter 3 & (d) Our Filter \\
	\end{tabular} 
	\caption{Comparison of filtered value of $\alpha$ with traffic states calculated using three other na{\"i}ve filters filters}
	\label{fig:naive-filter}
\end{figure}

Naive filter 1 (mean filter):
$
\mu_i  = \dfrac{1}{w}\sum_{j=i-w}^{i-1}y_i,~~
 y^f_i   = \dfrac{y_i - \mu_i}{\mu_i}
$

Naive filter 2 (simple smoothing): $
 y^f_i  = \dfrac{y_i - y_{i-1}}{y_{i-1}}
$

Naive filter 3 (quantile filter):
\begin{align*}
Y_i &= \{y_{i-w},...,y_{i-1}\}\\
Q_i& :  = Pr(Y_i \le y_i) = 0.5 ~ \mbox{(median)}\\
 y^f_i & = \dfrac{y_i - Q_i}{Q_i}
\end{align*}

The results of numerical experiments illustrate the following features. The filtered speed plot follows the measurement plot, which is an expected result. We have chosen a well-behaving sensor for the study and there is no outliers in the measurements. The filtering algorithm properly identifies  rate of change during break down and recovery regime. We can see that breakdown happens faster. This is a well known fact, that it takes less time for a queue to build up than to dissipate. This difference in time can be explained by driver's behavior and the fact that vehicles acceleration rate is lower then deceleration rate. However, the results for filtered probabilities of the switching variable $\alpha_t$ are less intuitive. Our filter properly identifies free flow regime in the morning, but then gets ``slightly confused'' during morning rush hour by assigning very close probabilities to recovery and breakdown regimes. We interpret this as saying that the Markov switching process model used to model $P(\alpha_t | \alpha_{t+1}, Z_t)$ is misspecified and need to be refined. This is a topic for our future research.

\section{Discussion}\label{sec:discussion}
We propose a mixture state-space model together with a particle filter and learning for tracking the state of traffic flow and other hidden variables such as flow regime and rate of change in the traffic flow speed. The proposed method is  flexible, in a sense that it does not require the state-space model to be Gaussian. Our approach does not rely on blind particle proposal for estimating the forecast distribution, but instead draws are taken from a smoothed distribution, which takes the measurement into consideration. Thus, our filter is fully adapted with exact samples from the filtering distribution are drawn. We used the sufficient statistics representation for state particles lead to a computationally efficient method. We formulate the traffic flow evolution equation as a hierarchal Bayesian model that is capable of capturing traffic flow discontinuities. 

Although we have focused on representation of state-space dynamics using multi-process DLM, the same approach will work for the kinematic wave theory based approach. When the evolution equation is given by the classical macroscopic traffic flow model, namely Lighthill-Whitham-Richards (LWR) partial differential equation \trbcite{lighthill1955kinematic, richards1956shock}. A Bayesian analysis of traffic flows using the LWR model is described in \trbcite{polson2014bayesian}. However, the analysis based on the  LWR model requires an estimation of conditions for a road segment, which is not always available. The statistical model described in here has the flexibility to avoid assumptions on locations of sensors. 

While we demonstrated the solution to the filtering problem in this paper, the fact that we provided a closed form solution to the filtering density and can do exact sampling from the filtering density allows us to use the same approach for particle learning and smoothing \trbcite{car10}. It works by augmenting the particle space to $\{x_t,s_t\}$, where $s_t$ is the sufficient statistic for the parameter space $\gamma$, i.e. $p(\gamma|x^t,y^t) = p(\gamma|\phi_t)$. In the DLM setting we can learn $\beta$, by removing the evolution equation for this parameter, rather then learning it directly from data, without relying on any model. Further research into predictive performance of such models is warranted.
\appendix
\section{Particle Filtering and Learning Methods}\label{app:pf}
Particle filtering and learning methods are designed to provide state and parameter inference via the set of joint posterior filtering distribution obtained in an on-line fashion [\trbcite{gordon1993novel}, \trbcite{carpenter1999improved},\trbcite{pitt1999filtering}, \trbcite{storvik2002particle}, \trbcite{car10}]. 

Let $y_t$ denote the data, and $\theta_t$ the state variable, in our context use $(x_t, \alpha_t)$.  Let $\phi$ denote the unknown parameters. For the moment, we suppress the conditioning on the parameters $\phi$. We will show how to update state variables and sufficient statistics for $\phi$.  First, 
we factorize  the joint posterior distribution of the data and state variables both ways as
\begin{align*}
p(y_{t+1},\theta_{t+1}|\theta_t) &=p(y_{t+1}|\theta_{t+1})p(\theta_{t+1}|\theta_t)\\
&=p(y_{t+1}|\theta_t)\,p(\theta_{t+1}|\theta_t,y_{t+1}) 
\end{align*}

The goal is to obtain the new filtering distribution $p(\theta_{t+1}|y^{t+1})$ from the current 
$p(\theta_{t}|y^{t})$. A particle representation of the previous filtering distribution is a 
random histogram of draws. It is denoted by
$
p^N(\theta_t|y^{t})=1/N\sum_{i=1}^N\delta_{\theta_t^{(i)}},
$ 
where $ \delta $ is a Dirac measure. As the number of particles increases $N\rightarrow \infty$ the law of large numbers guarantees that this distribution converges to the true filtered distribution $p(\theta_{t}|y^{t})$.

In order to provide random draws of the next distribution, we first resample $\theta_t$'s using the smoothing distribution 
$$
p(\theta_t|y^{t+1})\propto p(y_{t+1}|\theta_t)p(\theta_t|y^{t})
$$ 
obtained by  Bayes rule. Thus, we draw $\theta^{k(i)}_t$ by drawing the
index $k(i)$ from a multinomial distribution with weights
$$
w^{(i)}_t=\frac{p(y_{t+1}|\theta_t^{(i)})}{\sum_{j=1}^Np(y_{t+1}|%
\theta_t^{(j)})}.$$
We set $\theta^{(i)}_t=\theta^{k(i)}_t$ and {\em ``propagate''} to the next time period $t+1$ using the predictive
$$p(\theta_{t+1}|y_{1:t+1})=\int
p(\theta_{t+1}|\theta_t,y_{t+1})p(\theta_t|y_{1:t+1})d\theta_t.
$$
Given a particle approximation $ \{ \theta^{(i)} : 1 \leq i \leq N \} $ to
$p^{N}\left(  \theta_t|y^{t}\right)  $, we can use Bayes rule to write
\begin{align*}
p^{N}\left(  \theta_{t+1}|y^{t+1}\right)   &  \propto\sum_{i=1}^{N}p\left(
y_{t+1}|\theta_t^{\left(  i\right)  }\right)  p\left(  \theta_{t+1}|\theta_t^{\left(
i\right)  },y_{t+1}\right) \label{Mixture2}\\
&  =\sum_{i=1}^{N}w_{t}^{\left(  i\right)  }p\left(  \theta_{t+1}|\theta_t^{\left(
i\right)  },y_{t+1}\right)  \text{,}%
\end{align*}
where the particle weights are given by
\[
w_{t}^{\left(  i\right)  }=\frac{p\left(  y_{t+1}|\theta_t^{\left(  i\right)
}\right)  }{\sum_{i=1}^{N}p\left(  y_{t+1}|\theta_t^{\left(  i\right)  }\right)
}\text{.}%
\]
This mixture distribution representation leads to a simple simulation approach for propagating particles to the next filtering distribution. 

The algorithm consists of two steps:
\begin{align*}
&  \text{Step 1. (Resample) Draw }\theta_t^{\left(  i\right)  }\sim
Mult_{N}\left(  w_{t}^{\left(  1\right)  },...,w_{t}^{\left(  N\right)
}\right)  \text{ for }i=1,...,N\\
\text{ }  &  \text{Step 2. (Propagate) Draw }\theta_{t+1}^{\left(  i\right)  }\sim
p\left(  \theta_{t+1}|\theta_t^{\left(  i\right)  },y_{t+1}\right)  \text{ for
}i=1,...,N.
\end{align*}
 
To implement this algorithm, we need the predictive likelihood for the next observation, $y_{t+1}$, 
given the current state variable $ \theta_t $. It is  defined by
\[
p\left(  y_{t+1}|\theta_t\right)  =\int p\left(  y_{t+1}|\theta_{t+1}\right)  p\left(
\theta_{t+1}|\theta_t\right)  d\theta_{t+1}.%
\]
We also need the  conditional posterior for the next states $ \theta_{t+1} $ given $ (\theta_t , y_{t+1} )$.
It is  given by
\[
p\left(  \theta_{t+1}|\theta_t,y_{t+1}\right)  \propto p\left(  y_{t+1}|\theta_{t+1}%
\right)  p\left(  \theta_{t+1}|\theta_t\right) \; .
\]

This algorithm has several practical advantages. First,  it does not suffer from the problem of particle 
degeneracy which plagues the standard sample-importance resample filtering algorithms. This effect is 
heightened when $y_{t+1}$ is an outlier. Second, it can easily be extended to incorporate sequential 
parameter learning. It is common to also require learning about other unknown static parameters, denoted by 
$\phi$. To do this, we assume that there exists a conditional sufficient statistic $s_t$ for $\phi$
at time$\,t$, namely
$$
p( \phi | \theta_{1:t} , y_{1:t} ) = p( \phi | s_t )
$$
where $ s_t = s(  \theta_{1:t} , y_{1:t} ) $.
Moreover, we can propagate these sufficient statistics by the deterministic recursion $s_{t+1}=S(s_t,\theta_{t+1},y_{t+1})$, given particles $(\theta_t,\phi,s_t)^{(i)},$ $i=1,\ldots,N$. First,  we
resample $(\theta_t,\phi,s_t)^{k(i)}$ with weights
proportional to $p(y_{t+1}| ( \theta_t,\phi)^{k(i)})$. Then we propagate to the next 
filtering distribution
$p(\theta_{t+1}|y_{1:t+1})$ by drawing $\theta^{(i)}_{t+1}\,$from
$p(\theta_{t+1}|\theta^{k(i)}_t,\phi^{k(i)},y_{t+1}),\,i=1,\ldots,N$. We next
update the sufficient statistic for $i=1,\ldots,N$,
$$
s_{t+1}=S(s_t^{k(i)},\theta^{(i)}_{t+1},y_{t+1}).
$$ 
This represents a deterministic propagation. Parameter learning is
completed by drawing $\phi^{(i)}$ using $p(\phi|s^{(i)}_{t+1})$ for
$i=1,\ldots,N$. We now track the state, $\theta_t$, and conditional sufficient statistics, $s_t$, which will be used to perform off-line learning for $\phi$.

The algorithm now consists of four steps:
\begin{align*}
&  \text{Step 1. (Resample) Draw Index } k_t(i) \sim
Mult_{N}\left(  w_{t}^{\left(  1\right)  },...,w_{t}^{\left(  N\right)
}\right)  \text{ for }i=1,...,N\\
& \text{The weights are proportional to } p( y_{t+1} | ( \theta_t , s_t )^{(i)} ) \\
\text{ }  &  \text{Step 2. (Propagate) Draw }\theta_{t+1}^{\left(  i\right)  }\sim
p\left(  \theta_{t+1}| (\theta_t , s_t)^{k \left(  i\right)  },y_{t+1}\right)  \text{ for
}i=1,...,N.\\
\text{ }  &  \text{Step 3. (Update) Deterministic } s_{t+1}^{\left(  i\right)  }=
S\left( s_t^{k(i)} , \theta_{t+1}^{(i)}  ,y_{t+1}\right)  \text{ for
}i=1,...,N.\\
\text{ }  &  \text{Step 4. (Learning) Offline } \phi^{\left(  i\right)  }\sim
p\left(  \phi | s_{t+1}^{(i)} \right)  \text{ for
}i=1,...,N.\end{align*}
\bibliographystyle{trb}
\bibliography{ref}
\end{document}